*Tailwind turbulence: a bound on the energy available from turbulence for transit, tested in Kraichnan's model*

S. A. Bollt[1] and G. P. Bewley[2]
[1]*Department of Aerospace Engineering, California Institute of Technology, Pasadena, CA, USA*
[2]*Sibley School of Mechanical and Aerospace Engineering, Cornell University, Ithaca, NY, USA*

We investigate the unconstrained minimum energy required for vehicles to move through turbulence. We restrict our study to vehicles that interact with their environment through thrust, weight and drag forces, such as rotorcraft or submersibles. For such vehicles, theory predicts an optimum ratio between vehicle velocity and a characteristic velocity of the turbulence. The energy required for transit can be substantially smaller than what is required to move through quiescent fluid. We describe a simple picture for how a flight trajectory could preferentially put vehicles in tailwinds rather than headwinds, predicated on the organization of turbulence around vortices. This leads to an analytical parameter-free lower bound on the energy required to traverse a turbulent flow. We test this bound by computationally optimizing trajectories in Kraichnan's model of turbulence, and find that the energy required by point-models of vehicles is slightly larger than but close to our bound. Finally, we predict the existence of an optimum level of turbulence for which power is minimized, so that turbulence can be both too strong or too weak to be useful. This work strengthens previous findings that environmental turbulence can always reduce energy use. Thus, favorable trajectories are available to maneuverable vehicles if they have sufficient knowledge of the flow and computational resources for path planning.

**Introduction**

Turbulence is not noise [1,2]. It contains correlations. By virtue of these correlations we may extract work from it [3]. But how much? In devising new flight strategies, it is useful to know the upper bound on the advantage that might be gained. The success of a flight strategy can then be measured, in part, by its distance from this bound. If the turbulence is weak and the advantage is small, other contributions to the energetics may be more important than turbulence, such as steady drag. On the other hand, if turbulence is so strong that working against it is impossible, strategies that make use of turbulence may be necessary.

In sufficiently strong turbulence, fluctuations eventually overwhelm any control method whose primary aim is to suppress accelerations or disturbances. For insect-sized vehicles [*e.g.* 4], the condition of overwhelming turbulence can be the one that always prevails, since the atmosphere and oceans are always in motion and turbulent [5]. This scenario is also common for small autonomous underwater vehicles (AUV's), among others [*e.g.* 6]. An improved understanding of the energetics of travel through turbulence has widespread applications for emerging autonomous vehicle technologies, both flying and floating. It may also facilitate our understanding of the behavior of natural fliers and swimmers.

Instead of trying to suppress the effects of turbulent fluctuations, we consider trajectories that make use of them to travel faster or to use less energy. Such strategies invariably increase rather than decrease the accelerations of a vehicle as it follows favorable trajectories through the flow [3]. When there are no humans aboard to disturb, these accelerations may be large before they are consequential. By providing unlimited knowledge of the flow, unlimited computer resources, and unlimited control authority to our optimization problem, we provide a benchmark or upper bound for what such trajectories can theoretically achieve. The bound is useful for evaluating the potential of new strategies whose knowledge of the flow, computational resources, or actuation is limited. As new classical and machine learning techniques emerge for flow inference [7,8] and path optimization [3,9,10] one would expect the performance of real vehicles to approach but never reach the limiting performance we predict.

In this paper we introduce an analytic bound for flight in turbulence, which is a simple interpretation of the physics of vehicle transport in a flow field that represents an upper bound for the energetic advantage that turbulence can supply to a vehicle. The bound corresponds to a flight path that continually finds tailwinds without effort. We call it Tailwind Turbulence ("TW"). We expect the bound to be useful when vehicles are much smaller than the energetic scales of environmental turbulence, for instance for vehicles in the atmosphere that are much smaller than hundreds of meters. The bound is useful to evaluate the efficacy of new path planning strategies and their potential in applications. Finally, we argue that a simple physical picture for how the bound becomes too optimistic in high turbulence levels leads to the prediction that there exists an optimal level of turbulence from the point of view of the energetics of transit, even for vehicles with unlimited maneuverability.

Using numerical path optimization, we compare the TW bound to the performance of real trajectories found using unlimited flow knowledge, unlimited control authority, and nearly unlimited computational resources on a two dimensional (2D) turbulence model due to Kraichnan [11]. The Kraichnan model is a large step beyond random noise since the flow is correlated and incompressible, and it has been used to demonstrate basic theoretical results for passive scalars and for inertial particles that hold in real turbulence [21,16]. An additional advantage of the model is the speed at which we can compute ensemble averages from many different realizations of the flow relative to direct numerical simulations of turbulence. The model's disadvantages include weaker correlations and a narrower range of active scales relative to environmental turbulence. We use an approximation for the energy consumed by flight vehicles introduced in Bollt and Bewley [3] since its limiting forms have useful physical interpretations and because it facilitates the computer optimization of flight strategies. Relative to Bollt and Bewley [3], the approach we use here and the calculations we present are new and distinct: we study the properties of parameterized trajectories, rather than the properties of trajectories generated by certain dynamics. Here we also generalize the derivations to include all hydrodynamic vehicles that only experience thrust and drag, both following power laws, and weight. The framework makes it possible to compare straight flight through quiescent fluid ("QF") with the favorable trajectories that inertial particles generate naturally due to preferential sweeping by turbulence, which is called fast tracking ("FT") [15].

**Tailwind Turbulence Theory**

*i.* Definitions

We seek to minimize the energy, $E$, required for a trip by varying the path itself and the speed along the path. We shall see that this minimum depends on the level of turbulence relative to characteristics of the vehicle. We hold the vehicle mass $m$, and gravitational acceleration $g$ constant. The constant mass assumption applies generally to electric vehicles or to fuel-consuming vehicles on short trips. For long trips by vehicles that consume fuel, the analysis below needs to be adapted to account for changing mass.

We first consider the power required for travel, which depends on the way a vehicle generates motion. We analyze vehicles for which power is a function only of the force, $\bar{F}_T$, exerted by the vehicle on the fluid, such as rotorcraft, where the over bar denotes a vector. The dimensionless power for such vehicles is $\mathbb{P} = (\bar{\mathbb{F}}_T^2)^{p+1/2}$, where $\bar{\mathbb{F}}_T \equiv \bar{F}_T/mg$ is the dimensionless force, and the exponent $p$ describes the nonlinearity in the energy consumption. For rotorcraft the velocity induced by the propellers is usually high compared with the vehicle's velocity through the fluid, and actuator disc theory predicts that $p = 1/4$ [12]. For many vehicles under normal operating conditions, $p \in [0, 1/4]$, which is a narrow range, but cases such as propellor-driven airplanes where the force depends on the vehicle velocity relative to the fluid need to be treated separately.

To evaluate the force exerted by the vehicle on the fluid, we use a point model of the vehicle. This is a useful approximation for vehicles that are much smaller than the energetic scales in the flow [3]. In our model, forces are due to weight and buoyancy as well as thrust and drag. We allow thrust to point in any direction at any given time, which is a good model for real behavior if the vehicle can reorient itself much faster than the energetic structures in the flow evolve. The inertia of the vehicle includes the vehicle mass and added mass of fluid dragged along by the vehicle [22]. To simplify our analysis, we consider motion in the plane perpendicular to the acceleration of gravity only (Fig. 1), which eliminates the possibility of exchanges between kinetic and gravitational potential energy and focuses our analysis on those advantages that may arise purely from interactions with turbulent wind currents. This requires a component of thrust always to balance the (buoyant) weight of the vehicle.

According to Newton's second law of motion, the thrust force required to follow a specified path is the difference between the inertia of the vehicle and the sum of the drag and body forces. Neglecting added mass and buoyancy for brevity, $\bar{F}_T = m\bar{a}' - \bar{F}_d - m\bar{g}$, where $\bar{a}' = d\bar{u}'/dt'$ is the vehicle acceleration, $\bar{u}' = \bar{u}'(t')$ is the vehicle velocity, and the prime denotes aspects of the vehicle's state with dimensions, or units. Term by term, our equation for the thrust force is

$$\bar{F}_T = \mathbb{A} m \frac{d\bar{u}'}{dt'} - \frac{m}{\tau_d W^{d-1}} |\bar{w}' - \bar{u}'|_2^{d-1}(\bar{w}' - \bar{u}') - \mathbb{B} m\bar{g}, \quad (1)$$

where the drag is proportional to $\bar{w}' - \bar{u}'$ and is nonlinear if $d \neq 1$; $\bar{w}' = \bar{w}'(\bar{x}', t')$ is the wind velocity field, which has zero mean: $\langle \bar{w}' \rangle = 0$ (brackets, $\langle \cdot \rangle$, henceforth denote time averaging). The vehicle's drag constant is represented by the timescale, $\tau_d$; $\tau_d$ measures the initial slope of the vehicle's return to equilibrium from a step-gust of strength $W$. There are two dimensionless parameters here. The parameter $\mathbb{A} \equiv (m + m_{add})/m$ describes the effect of the mass, $m_{add}$, of fluid that accelerates with the vehicle. Dense vehicles have a value of $\mathbb{A}$ close to 1, and neutrally buoyant vehicles have a value as large as 3/2. The second parameter, $\mathbb{B} \equiv (m - m_f)/m$, describes the effect of buoyancy, where $m_f$ is the mass of the fluid displaced by the vehicle; $\mathbb{B}$ is also close to 1 for heavy vehicles, and is zero for balloons or submarines. We will in this paper focus on the case that $\mathbb{A} \approx \mathbb{B} \approx 1$, but we retain them here as variables to enable consideration of how our results change with departures from this case, and for neutrally buoyant vehicles in particular.

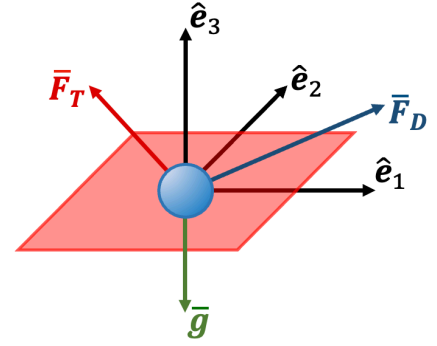

**Figure 1**: The coordinate system for the point vehicle (blue). Gravity ($\bar{g}$) points downward, drag ($\bar{F}_D$) points opposite to the direction of motion, and thrust ($\bar{F}_T$) can be in any direction. Motion is restricted to the plane normal to gravity (red) and is in the direction of $-\hat{e}_2$ on average.

In dimensionless form, the thrust force given in Eq. 1 is

$$\bar{\mathbb{F}}_T = \frac{1}{\mathbb{G}} \left[ \mathbb{A} St \frac{d\bar{u}}{dt} - |\bar{w} - \bar{u}|_2^{d-1}(\bar{w} - \bar{u}) \right] - \mathbb{B}\hat{g}, \quad (2)$$

where $\bar{u}(\bar{x}, t) = \bar{u}'/W$ and $\bar{w}(\bar{x}, t) = \bar{w}'/W$; $\bar{x} = \bar{x}'/L$ and $\bar{t} = \bar{t}'W/L$; $W$ is the typical amplitude of the turbulent wind speed fluctuations so that the mean kinetic energy density is $(1/2)\rho W^2$; $L$ is the correlation length of the turbulence; and $\hat{g}$ is the unit vector in the direction of gravity. Two additional dimensionless parameters emerge when making the force dimensionless [3]: $\mathbb{G} \equiv g\tau_d/W$ describes the turbulence intensity and decreases to 0 for strong turbulence ($W \to \infty$) and for vehicles that are tightly coupled to the flow ($\tau_d \to 0$), where $\tau_d$ is the characteristic response time of the vehicle to changes in wind speed ($\tau_d$ is small for lightweight vehicles with large drag); $St \equiv \tau_d W/L$ compares the vehicle response time with the turbulence timescale, $L/W$, so that $St$ is small for vehicles that are tightly coupled to spatially extended and slowly evolving turbulence. The Reynolds number is not explicitly a parameter in this problem, though the Reynolds number of the vehicle in part determines the appropriate choice of the drag exponent, $d$.

We now write the dimensionless energy in terms of the dimensionless mean vehicle speed in the direction of desired motion, $\mathbb{U} \equiv d_f/t_f W$: $\mathbb{E} \equiv \mathbb{G}\langle \mathbb{P}\rangle/\mathbb{U}$. This definition of energy is equivalent to the cost of transport (COT), $E/mgd_f$, up to a constant factor (the factor is intrinsic to the vehicle's design and is related to thrust-to-power ratio, and drag). We then substitute our expression for the power in terms of the thrust, and write the average power explicitly as a time integral, so that

$$\mathbb{E} = \frac{\mathbb{G}\mathbb{B}^{2p+1}}{\mathbb{U}} \frac{1}{t_f} \int_0^{t_f} \left| \frac{1}{\mathbb{G}\mathbb{B}} \left[ \mathbb{A} St \frac{d\bar{u}}{dt} - |\bar{w} - \bar{u}|_2^{d-1}(\bar{w} - \bar{u}) \right] - \hat{g} \right|_2^{2p+1} dt . \quad (3)$$

In this model, the key dimensionless parameters that control the energetic cost of point vehicles in a turbulent fluid are $p, d, \mathbb{B}, \mathbb{G}$, and the product $\mathbb{A} St$, with $\mathbb{U}$ being an optimization variable and not a parameter. We will show that $\mathbb{A} St$ appears not to strongly affect optimal energy consumption in our simulations, especially for small vehicles in atmospheric turbulence, and that a more natural grouping of parameters may exist. If we ignore the inertial term of Eq. 2 we find that energy use is controlled by weight and drag only, with the former being a constant. The cost of working against gravity may be augmented or entirely replaced in certain applications by a resting energy or basal metabolic cost, with no impact on our conclusions.

*ii. Quiescent flow*

In this section we describe useful reference cases that follow classical aerodynamics theory expressed in our notation [12]. For straight line trips at constant velocity, for instance, $\bar{u} = [0 \ \mathbb{U} \ 0]$, which defines the dimensionless vehicle velocity in terms of its speed toward its destination relative to the typical fluctuations in the wind velocity, $\mathbb{U} = U/W$. Since the acceleration is zero, the force, power, and energy do not depend on $St$. Our simulations, described below, show this to be a good approximation even for optimized and curved trajectories through turbulence. For straight trips, in any case, the energy does not depend on the path and the problem reduces to determining whether $dE/dU = 0$ for a nontrivial flight speed $U^*$.

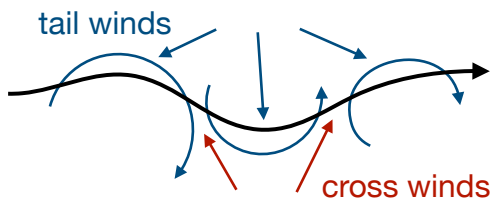

**Figure 2**: A favorable trajectory (black curve) through turbulence (blue curves) finds tailwinds more often than headwinds, but may pass through regions with cross winds in order to do so.

When there is no turbulence (when $\bar{w} = 0$ everywhere), the dimensionless energy required for transit is $\mathbb{E}_{QF} \equiv \left(\mathbb{U}^{2d} + \mathbb{G}^2\mathbb{B}^2\right)^{p+1/2}/\left(\mathbb{U}\mathbb{G}^{2p}\right)$, using "QF" to denote quiescent flow. Observe that the energy required to fly through quiescent flow, $\mathbb{E}_{QF}$, has a minimum in terms of the flight speed, $\mathbb{U}$. The minimum arises from the competition between the energy required to work against gravity and the energy required to work against drag. For linear drag and typical rotorcraft ($d = 1$ and $p = 1/4$), the minimum energy is achieved for flight velocities given by $\mathbb{U}^*_{QF} = \sqrt{2}\mathbb{G}\mathbb{B}$. Thus the flight speed required to minimize energy in the absence of turbulence is fixed. The implicit appearance of the turbulent wind speed, $W = 0$, as the reference velocity in the dimensionless parameters can be disregarded since it cancels out. For this reason it is convenient to define $\mathbb{V} \equiv \mathbb{U}/\mathbb{G}$ since its optimized value, $\mathbb{V}^*$, is well defined when $\mathbb{G} \to \infty$. This has the further convenience that the energy can be calculated simply from $\mathbb{E} = \langle \mathbb{P} \rangle / \mathbb{V}$. The corresponding minimum energy, $\mathbb{E}^*_{QF}$, depends only on $\mathbb{B}$ (as $\mathbb{B}^{1/2}$ for $d = 1$ and $p = 1/4$), which can be checked by substitution of $\mathbb{U}^*_{QF}$ into $\mathbb{E}_{QF}$. Finally, since the COT is the same as $\mathbb{E}$ except for a multiplicative constant (see section *i*), comparing $\mathbb{E}_{QF}$ to the energy (measured in Joules, for instance) required to move the vehicle's weight a given distance through a quiescent flow allows conversion between $\mathbb{E}$, $E$, and COT for all of the scenarios we explore. The same is true for the dimensionless power, $\mathbb{P}$.

*iii.* Tailwind Turbulence

We now move beyond classical theory. To minimize energy in turbulence, we need to calculate the force exerted by the vehicle along a trajectory through a spatially evolving flow field. The challenge in doing so efficiently, for vehicles of the type we consider, is to find a path that puts a vehicle in tailwinds more often than headwinds. This must be done without expending too much additional energy to change the path, otherwise the potential benefit is lost. A simple picture of turbulence is that it is composed of vortical structures that form cells connected by saddle points [13]. An efficient way to traverse the flow is to follow streamlines within each cell, and then to hop across saddle points from one cell to the next, building a nearly optimal trajectory as you go (Fig. 2). In this picture, the vehicle experiences tailwinds on the scale of the amplitude of the velocity fluctuations, $W$, along most of its trajectory. Furthermore, the cells may be fortuitously aligned and organized so that the length of the trajectory is that of a line between the origin and destination. When turbulence is not too strong, the vehicle can make forceful adjustments to its path, and the path length will be equal to the length of a line up to a ($\mathbb{G}$-dependent) factor on the order of the sum of two sides of a square to its diagonal ($\sqrt{2}$). We neglect this factor since we expect it to be small in comparison with other contributions to the energy until $\mathbb{G}$ is very small. With vanishing probability on long trips, it may be possible for the vehicle to find a path that suffers no crosswinds at all (a path that is also a path-line of the flow). Otherwise, the crosswinds scale with $W$, meaning that they diverge toward large values in strong turbulence. We neglect crosswinds for now, but revisit them in the discussion (below). The effect of path length and cross winds, we expect, make

Tailwind Turbulence cease to be a useful measure for achievable advantages in strong turbulence (for $\mathbb{G} \to 0$).

We define Tailwind Turbulence ("TW") as a straight path along which turbulence blows in the direction of flight without the need for extra control input, so that $\bar{w} = [0\ 1\ 0]$. In other words, we find paths with constant tailwinds of strength equal to the turbulence velocity, $W$. As in quiescent flow, the vehicle velocity is approximately constant so that there are negligible accelerations and $\bar{u} = [0\ \mathbb{U}\ 0]$. The dimensionless energy in the TW case is

$$\mathbb{E}_{TW} \equiv \frac{1}{\mathbb{U}\mathbb{G}^{2p}} \left([1-\mathbb{U}]^{2d} + \mathbb{G}^2\mathbb{B}^2\right)^{p+1/2}. \tag{4}$$

The same result holds for vehicles with negligible inertia ($St \to 0$), moving along curved paths. As for the quiescent flow case, the energy contains a minimum in the sense that $d\mathbb{E}_{TW}/d\mathbb{U} = 0$ for a certain vehicle speed: $\mathbb{U}^*_{TW} = (1/2)(\sqrt{8\mathbb{G}^2\mathbb{B}^2+9} - 1)$ for linear drag and typical rotorcraft ($d=1$ and $p=1/4$). For nonlinear drag and arbitrary power nonlinearities, general expressions can be found computationally, as can the corresponding energies. For linear drag and typical rotorcraft, the lower bound on the energy required for flight through turbulence is

$$\mathbb{E}^*_{TW} = \frac{2^{1/4}}{\mathbb{G}^{1/2}} \frac{\left[9 - 3\sqrt{8\mathbb{G}^2\mathbb{B}^2+9} + 6\mathbb{G}^2\mathbb{B}^2\right]^{3/4}}{\sqrt{8\mathbb{G}^2\mathbb{B}^2+9} - 1}. \tag{5}$$

For weak turbulence ($\mathbb{G} \to \infty$), the vehicle velocity, $\mathbb{U}^*_{TW}$, grows in proportion to $\mathbb{G}$, meaning that the optimum velocity approaches a constant equal to the quiescent flow optimum, and transit requires the quiescent flow energy as expected. Though we do not expect Tailwind Turbulence to be a useful lower bound in strong turbulence ($\mathbb{G} \to 0$), in this limit the vehicle velocity approaches a constant as $\mathbb{U}^*_{TW} \to 1 + (2/3)\mathbb{G}^2\mathbb{B}^2$, meaning that the optimum vehicle velocity approaches the fluctuating turbulence velocity, $W$, quadratically from above. Furthermore, the energy vanishes to zero as $\mathbb{E}^*_{TW} \to \mathbb{G}$, so that under the Tailwind Turbulence assumption trips require no energy in the limit of strong turbulence. In essence, the small $\mathbb{G}$ limit represents the vehicle behaving as a passive tracer of the flow whose trajectory happens to be straight and to bring it to its desired destination. Since these flow trajectories are unlikely, we conclude that the model will not be useful if $\mathbb{G}$ is too low. This issue is addressed in the discussion section. If compared with disturbance rejection [3], which requires power diverging toward infinity for small $\mathbb{G}$, the energy savings grow arbitrarily large toward small $\mathbb{G}$.

**Methods**

Here we describe how we parameterize the trajectories and optimize the parameters to test the tightness of the lower bound given by TW. We compare the analytic bound described above with simulations performed on a point-mass traveling through a deterministic flow. All flight path optimizations and statistics were computed over 100 flows with initial conditions selected at random.

The flow we chose is the incompressible two-dimensional (2D) model of statistically stationary, homogeneous and isotropic turbulence described in Refs. [3,11,14]. It is not a direct numerical simulation of the equations of motion, and it may underestimate the correlations in turbulence beneficial to transit through turbulence [3,15]. An important distinction between this 2D random flow field and 2D turbulence is that the random flow field does not contain a reverse energy cascade. We chose the Kraichnan model over DNS data for this study due to computational limitations. This is due to the exhaustive nature of the global optimizations we performed, and the fact that each point in parameter space represents an ensemble average — each point in subsequent figures represents an evaluation of the flow velocity at hundreds of millions, to billions, of points. The Kraichnan model is not constrained by the momentum equation, but it is incompressible. This flow field is preferred over simpler flows such as the cellular flow field because it is multi-scale in space and time, it has finite temporal correlation time, it has a well defined maximum energy containing scale, and it is periodic on length and time scales so large as to be essentially aperiodic. These properties prevent simple strategies from being unrealistically effective, and also appears to produce ergodicity in passive particle dynamics, which makes the sampling problem much more straightforward.

We represent flight paths, $x(t)$, between $x(0)$ and $x(t_f)$ as a Fourier series and optimize over its coefficients. That is, we let

$$\boldsymbol{x}(t) = \begin{bmatrix} a_0 \\ -\mathbb{U}t \end{bmatrix} + \begin{bmatrix} \sum_{i=1}^{n} a_i \sin \frac{\pi i t}{2 t_f} \\ \sum_{i=1}^{n} b_i \sin \frac{\pi i t}{t_f} \end{bmatrix}, \quad \sum_{i=1}^{\lfloor \frac{n+1}{2} \rfloor} (2i-1) b_{2i-1} = 0, \tag{6}$$

with the sine series automatically satisfying the initial condition. The initial position is $x(0) = [a_0, 0]$, and the average flight speed is $\mathbb{U}$ in the $-\hat{e}_2$ direction. To avoid producing trajectories whose energy is dominated by escape from a bad initial condition (most initial conditions chosen at random are sub-optimal because most points in state space are not part of an optimal trajectory), the initial position in the horizontal direction was free but bounded to ±5 units in the $\hat{e}_1$ direction. The value of 5 was chosen because it is small compared with the length of the trajectories and large compared with the size of flow structures (~1). Furthermore, we found that increasing or decreasing the tightness of this constraint did not affect our conclusions. Since the energy content of the flows is concentrated around fluctuations with a timescale of $L/U$ and with length scale $L$, we optimized over a finite number of modes, $n$, up to $n$ = 32. The restriction on odd terms of $b_i$ ensures the initial and final velocities in $\hat{e}_2$ are the same, which eliminates paths that convert initial kinetic energy into progress toward the destination (as opposed to achieving the goal of finding less costly paths through the flow). We did not restrict the average flight speed in $\hat{e}_1$ or otherwise restrict its initial and final velocity. Differences between initial and final velocities orthogonal to the mean flight direction cannot be extracted as useful work since our model does not contain lift forces. The globally optimal trajectory is therefore approximated by a point (or set of points if it is not unique) in a $2n + 2$ dimensional parameter space $(\mathbb{U}, a_i, b_j), i \in 0,1,...n, j \in 1,2,...n$.

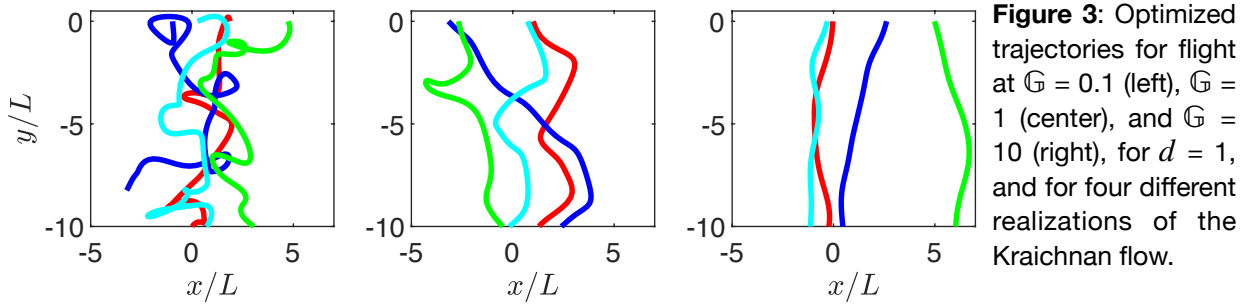

**Figure 3**: Optimized trajectories for flight at $\mathbb{G}$ = 0.1 (left), $\mathbb{G}$ = 1 (center), and $\mathbb{G}$ = 10 (right), for $d$ = 1, and for four different realizations of the Kraichnan flow.

Figure 3 shows examples of the trajectories we found by these methods. We found optimal trajectories through flows by optimizing the $2n + 2$ parameters using Matlab's global optimization function, which uses multiple initial guesses to seed parallel implementations of a local optimization method, followed by further searching in promising regions of parameter space, in an attempt to find a global minimum near or among the set of local minima discovered. We used gradient descent (*fmincon*) for the local scheme with an interior-point method for handling constraints, an optimality tolerance of $10^{-6}$, and both a step tolerance and a minimum difference tolerance of $10^{-5}$. There is no guarantee that the trajectories found are globally optimal — they are locally optimal. It seems likely to us that the paths were close to globally optimal because the variance in optimal energy between different random flows was much smaller than $\mathbb{E}_{QF}$, despite there being many local minima with $\mathbb{E} > \mathbb{E}_{QF}$. Also, we performed the optimization for certain selected trials for at least twice the iterations required for the energy to reach a minimum. It is possible in principle to guarantee that the global minimum energy path is found. For example, Dijkstra's algorithm could be used on a discretized state-space transition network of sufficient resolution. However the computational expense of producing the network and optimizing over it is beyond our computational resources. To estimate the dimensionless energy, $\mathbb{E}$, we averaged the energy over ensembles of trajectories of fixed length in time, $t_f$ (each ensemble point approximately minimizes the energy normalized by the distance that each trajectory covered during the simulation). Trajectories with a duration of $t_f = 10\pi/(1 + \mathbb{G})$ were sufficiently long for convergence of $\mathbb{E}$, but short enough to ensure that $n = 32$ would allow full exploitation of all the flow features. The global optimization was initialized using a straight-line path with $\mathbb{U} = \mathbb{U}^*_{QF}$. The constraints $10^{-3} \leq \mathbb{U} \leq 10\mathbb{G}$ and $|b_i|, |a_i| \leq 100$ sped up the optimizations by filtering out obviously bad trajectories.

**Results**

We find that the energy required for transit along optimized trajectories is always lower than the energy required in quiescent flow. The energy of optimized trajectories also decreases in stronger turbulence (smaller $\mathbb{G}$). Figure 4 shows the sample averaged minimum energy, $\mathbb{E}^*_{opt}$, normalized by $\mathbb{E}^*_{QF}$ for non-inertial ($St = 0$) rotorcraft for both linear ($d = 1$) and nonlinear ($d = 2$) drag as functions of $\mathbb{G}$. We find that Tailwind Turbulence bounds both cases ($d = 1$ and $d = 2$) as well as the energy for the fast-tracking ("FT") strategy discussed below. All uncertainty bars are 95% confidence intervals for the mean (under the assumption that each sample is taken from the same distribution).

We observe that a particular parameter grouping tends to collapse data generated with different drag nonlinearities ($d = 1$ and $d = 2$). We generate this grouping in the following way. In Eq. 3, the energy associated with producing a certain velocity relative to the fluid, $\Delta u$, scales with $|\Delta u|^d / (\mathbb{G}\mathbb{B})$ for large $\Delta u$. For a given flow, $|\Delta u|$ is fixed by the kinematics required to achieve a given trajectory. Therefore, the dimensionless parameter $\mathbb{S} \equiv (\mathbb{G}\mathbb{B})^{1/d}$ describes the degree of coupling between the flow and the vehicle since it is this term, rather than $\mathbb{G}$ alone, that determines the cost of moving relative to the flow. We expect the collapse of $(\mathbb{G}, \mathbb{B}) \to \mathbb{G}\mathbb{B}$ to be exact, and an approximate collapse with respect to the exponent, $d$. The pre-factor in Eq. 3, $\mathbb{G}\mathbb{B}^{2p+1}/\mathbb{U}$, is constant for a given vehicle and impacts energy consumption but not the optimal trajectory. This prefactor can be written in a simple way, $\mathbb{Q}/\mathbb{V}$, by defining $\mathbb{Q} \equiv \mathbb{B}^{2p+1}$. We then define a scaling for energy, $\mathbb{H} \equiv \mathbb{E}/\mathbb{Q}$, so that the energy is approximately a function of just two dimensionless parameters,
$\mathbb{E}(p, d, \mathbb{B}, \mathbb{G}, \mathbb{A}St, \mathbb{U}) \rightsquigarrow \mathbb{H}(\mathbb{S}, \mathbb{V})$. The optimal scaled energy is approximately a function of just one variable: $\mathbb{H}^* = \mathbb{H}^*(\mathbb{S})$.

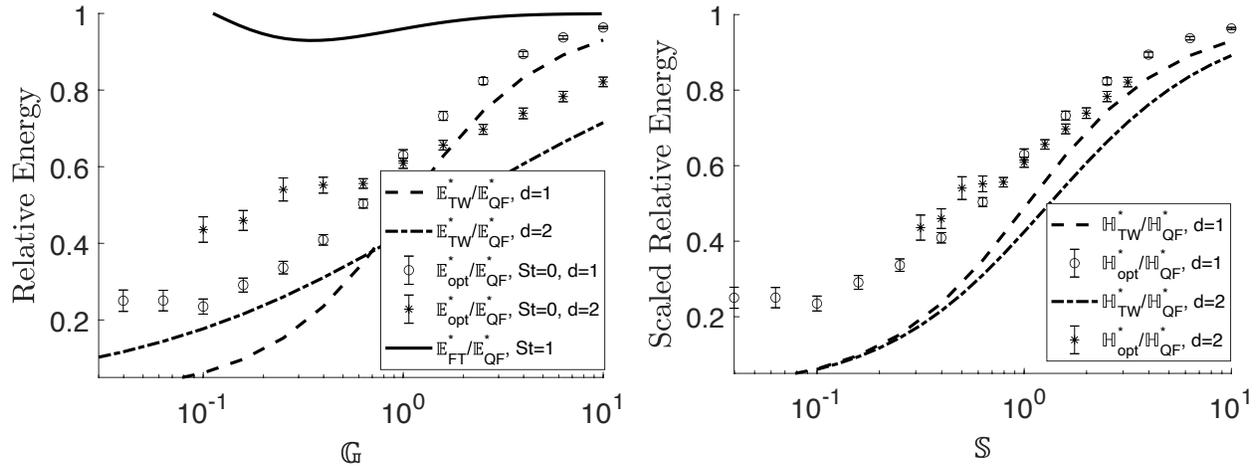

**Figure 4**: *Left:* The energy required to traverse the Kraichnan turbulence model ($\mathbb{E}^*_{opt}$, circles for linear drag and stars for quadratic drag), which we minimized by optimizing flight trajectories, decreases in stronger turbulence (smaller $\mathbb{G}$). The energy is bounded from below by Tailwind Turbulence ($\mathbb{E}^*_{TW}$, dashed lines: linear, dash-dot: quadratic drag, see Eq. 4). For reference, we include the energy required by "fast tracking" trajectories ($\mathbb{E}^*_{FT}$, solid lines) that vehicles generate automatically by responding only to local, instantaneous accelerations in a specific way (see Ref. [3]). *Right:* An approximate collapse with respect to changes in the nonlinearity in drag ($d = 1$ and $2$) is achieved when the data are plotted in terms of new parameters, $\mathbb{S}$ and $\mathbb{H}$, as described in the text.

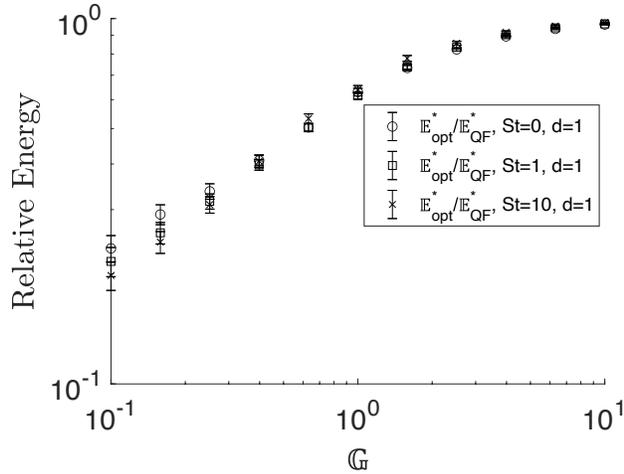

**Figure 5**: The optimized energy required to traverse turbulence ($\mathbb{E}^*_{opt}$) decreases in stronger turbulence (smaller $\mathbb{G}$) by a similar amount regardless of the inertia of the flight vehicle ($St$). For $St = 0$ (circles), the data are the same as in Fig. 1 and the vehicles are tightly coupled to the flow. For $St = 1$ (squares), a resonance between the time it takes for the flow to change and the time it takes for the vehicle to respond leads to fast tracking for unconstrained flight [3,15]. For $St = 10$ (crosses), vehicles are so heavy that they tend to be insensitive to turbulence and require large forces to change course.

In Fig. 5, we investigate the influence of vehicle inertia ($St$) on energy consumption for vehicles experiencing linear drag. It is somewhat surprising that the energy changes very little with large changes in $St$, and it suggests that working against drag dominates over work to accelerate a vehicle along optimal trajectories for any $\mathbb{G}$. Having large inertia ($St \gg 0$) appears to be detrimental for large $\mathbb{G}$ but it is slightly beneficial for small $\mathbb{G}$. The switch between these behaviors occurs at roughly $\mathbb{G} = 1$. The apparent unimportance of inertia to these globally optimized trajectories supports our neglect of vehicle accelerations along curved paths, and supports the idea that Tailwind Turbulence corresponds to a mapping of trajectories along curved path lines onto straight paths.

In Fig. 6, we break down the contributions to the dimensionless energy, $\mathbb{E} = \langle \mathbb{P} \rangle / \mathbb{V}$, which are the dimensionless average power, $\langle \mathbb{P} \rangle$, and the dimensionless average vehicle velocity in the direction of mean travel, $\mathbb{V}$. We discuss first the velocity. Figure 6 shows how the average velocity of optimal trajectories is close to the Tailwind Turbulence bound, which is especially true for the average velocity magnitude. The dimensionless average velocity in the mean direction of travel, $\mathbb{U}$, appears to be bounded from above by the value predicted by Tailwind Turbulence while the average dimensionless velocity magnitude, $\langle |\bar{u}| \rangle$, slightly exceeds the Tailwind Turbulence predictions for $\mathbb{G} = 1$ and slightly lags the Tailwind Turbulence predictions for $\mathbb{G} = 0.1$. The fact that the average velocity magnitude and the average velocity in the direction of desired travel increasingly differ as $\mathbb{G}$ shrinks is an indication that the optimal paths are becoming more tortuous. At the smallest values of $\mathbb{G}$ that we explored, the data begin to diverge from Tailwind Turbulence, which may be due to the increasing cost associated with jumps between favorable parts of the flow, as discussed in the next section. The optimized average velocities are close to the Tailwind Turbulence bound and the optimized energy is systematically larger, it must be the case that the optimized power is systematically larger than the Tailwind Turbulence bound. We discuss this next.

Figure 6 shows qualitative differences between the average power on optimal trajectories and the power predicted by Tailwind Turbulence. For $\mathbb{G} > 0.6$ the power required to follow optimal trajectories exceeds the Tailwind Turbulence prediction, but follows the same qualitative trend. For $\mathbb{G} < 0.6$ there is a change in average power usage. Tailwind Turbulence predicts that average power continues to fall as $\mathbb{G}$ decreases (until it reaches the power required to hold the vehicle aloft for $\mathbb{G} \to 0$). In contrast to this monotonic decrease toward $\mathbb{G} \to 0$, the average power of optimal trajectories rises as $\mathbb{G}$ decreases below 0.6. This indicates that different considerations are needed to construct a useful bound for energy consumption in strong turbulence (for $\mathbb{G} \to 0$). In this regime, we suspect that passive strategies will be more successful than forceful path optimization.

**Discussion**

Here we estimate the turbulence level, $\mathbb{G}_o$, below which Tailwind Turbulence may no longer be a useful measure of the energetic advantage of flying with turbulence. We anticipate that the breakdown of Tailwind Turbulence's utility at high turbulence levels ($\mathbb{G} \to 0$) can be understood in terms of the growing importance of crosswinds toward small $\mathbb{G}$. Crosswinds are a necessary consequence of generating a mean vehicle velocity in a turbulent flow with zero mean. It becomes necessary to jump between nearby path lines, and across increasingly strong winds at higher turbulence intensities. We model the influence of crosswinds by letting $\bar{w} = [c\ 1\ 0]$, where $c$ is a small parameter equal to zero for travel along path lines of the flow throughout the trip, and less than one for trips that have favorable path lines but in disconnected parts of the flow. Another interpretation is that $c$ is related to the angle between the flight velocity and the local flow velocity. The corresponding dimensionless power then contains an additional term, so that

$$\mathbb{E}_{TW,c} \equiv \frac{1}{\mathbb{U}\mathbb{G}^{2p}} \left[ ([1-\mathbb{U}]^2 + c^2)^d + \mathbb{G}^2 \mathbb{B}^2 \right]^{p+1/2}, \tag{7}$$

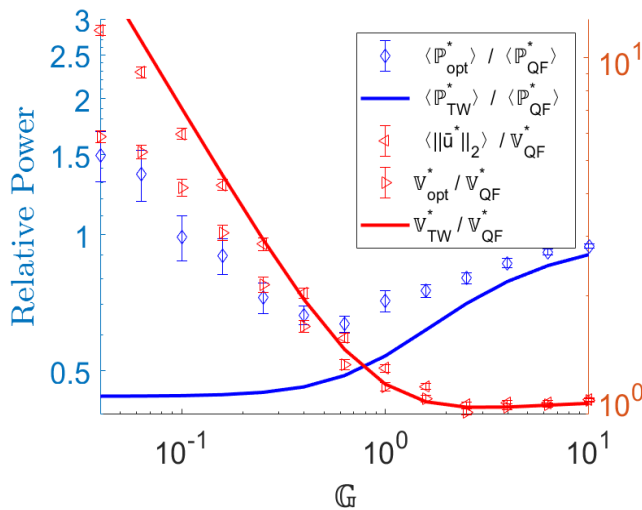

**Figure 6**: The average power generated along trajectories optimized to minimize energy consumption ($\mathbb{P}^*_{opt}$, blue diamonds) exhibits a minimum near $\mathbb{G} = 0.6$, which is not predicted by Tailwind Turbulence ($\mathbb{P}^*_{TW}$, blue curve), but can be explained by considering crosswinds that grow stronger toward smaller $\mathbb{G}$ as explained in the text. The vehicle's average speed (red triangles) matches the one predicted by Tailwind Turbulence ($\mathbb{V}^*_{TW}$, red curve) for all $\mathbb{G}$, meaning that flight vehicles on optimized paths move at the speed of turbulent fluctuations in strong turbulence.

to be compared with Eq. 4. Holding $c$ constant as a first approximation, since it cannot vanish for $\mathbb{G} \to 0$ if the vehicle makes progress toward its destination (since $\langle w \rangle = 0$) [16], we find that $\mathbb{E}_{TW,c}$ has a minimizing value of the flight speed, $\mathbb{U}^*_{TW,c}$, as did $\mathbb{E}_{TW}$. It is easier to analyze the case of linear drag and a square root power nonlinearity ($d = 1$ and $p = 1/2$), for which we find $\mathbb{U}^*_{TW,c} = \sqrt{1 + c^2 + \mathbb{G}^2 \mathbb{B}^2}$, which is close to one for small $c$ and $\mathbb{G}$ as expected.

We predict that for a given vehicle, there exists a turbulence level that minimizes energy consumption. That is, the optimum energy, $\mathbb{E}^*_{TW,c}(\mathbb{G})$, which is achieved at any given $\mathbb{G}$ by flying at the speed given by $\mathbb{U}^*_{TW,c}$, itself has a minimum in the sense that $d\mathbb{E}^*_{TW,c}/d\mathbb{G} = 0$ for a particular value of $\mathbb{G}$. In the case that $d = 1$, $p = 1/2$, and for small $c$, the optimum occurs near $\mathbb{G}_o \approx c/\mathbb{B}$. For turbulence weaker than the optimum ($\mathbb{G} \gg \mathbb{G}_o$), there simply is not enough to be gained from the fluctuations. For stronger turbulence ($\mathbb{G} \ll \mathbb{G}_o$), inevitable crosswinds along the trajectory require so much energy to work against that they outweigh the advantage gained by finding tailwinds. In such strong turbulence, the best strategy may be to head in the direction of the destination without attempting to account for turbulent disturbances, or to minimize the increased path length resulting from this strategy, rather than to find tailwinds. In practice, the value of $c$, or the normalized amplitude of the typical crosswinds, probably depends on the structure of turbulence and on specifics of the vehicle in a way worth exploring.

Our simulations do not show a clear minimum energy over the range of $\mathbb{G}$ that we computed. However, the power generated by a vehicle flying at the minimum energy speed, $\mathbb{U}^*_{TW,c}$, itself has a minimum near $\mathbb{G}_{Pmin} \approx \sqrt{2c}/\mathbb{B}$, if $c$ is small, $d = 1$ and $p = 1/2$. This minimum power is achieved at turbulence levels that are a factor of $\sqrt{2/c}$ *larger* than the level where our theory predicts the energy reaches its minimum ($\mathbb{G}_o$). Indeed, the data (Fig. 6) show a minimum in the power near 0.6 for $p = 1/4$, which suggests a value for $c$ smaller than 0.2. This value compares favorably with the root-mean-square of the angle between the vehicle velocity and the local wind velocity at small $\mathbb{G}$, which we found to be about 0.2 radians in our simulations. The corresponding energy minimum would then be near $\mathbb{G}_0 \approx 0.2$, which cannot be ruled out by our data, and requires simulations at smaller $\mathbb{G}$ to confirm.

The nearly globally optimal trajectories we found have significantly lower energy than the energy for the FT strategy found by Bollt and Bewley [3] on the same flows (Fig. 4). The FT strategy is a local rule for how to respond to vehicle accelerations inspired by the observation of increased settling velocities of particles in turbulence. It is interesting to ask how closely a local strategy can approach the global optima we present here. A substantial increase in settling speed enhancement is present in DNS generated isotropic turbulent flows [14,15] and in real turbulent experiments [15,17] over the 2D Kraichnan flow we used here [11]. This suggests the intriguing possibility that in real turbulence the gap in performance between the optimum strategy and the fast-tracking strategy is smaller than it is in our model turbulence.

We comment briefly on nearly neutrally buoyant vehicles, for which $\mathbb{B} \to 0$. According to our theory, they can produce net motion in a desired direction with vanishingly small energy regardless of the level of turbulence. It is well known that neutrally buoyant vehicles can achieve arbitrarily low cost of transport in quiescent flows by moving arbitrarily slowly (this holds true as long as there is no other energetic cost that integrates over time, associated with resting energy or basal metabolic costs for instance [15]). Perhaps surprisingly, this conclusion also holds for arbitrary levels of turbulence. This must be true because turbulence enhances the average speed of a passive particle with linear drag in a turbulent flow (the passive particle strategy is within the set of all possible strategies available to neutrally buoyant vehicles). Optimal trajectories for neutrally buoyant vehicles represent the $\mathbb{S} \to 0$ limit.

**Conclusions**

We showed that when information about a turbulent flow is available everywhere and for all times, optimized paths can be found that reduce energy consumption by up to a factor of four in simulations. The energy required for these trajectories is bounded by an analytic and parameter-free formula that we call Tailwind Turbulence, which depends on the strength of turbulence, normalized in a specific way. The energy bound decreases monotonically toward higher turbulence levels. Toward the highest turbulence levels in simulations, the optimized power increases away from the Tailwind Turbulence bound in a way consistent with our prediction of the existence of an optimal level of turbulence. We identify the cause as the increasing importance of cross winds in strong turbulence. For typical environmental turbulence, we expect the optimum turbulence level for extracting energy to lie somewhere between what insects and bacteria experience ($\tau_d = O(10^{-2})\,s$ or less, corresponding to $\mathbb{S} \lesssim 0.1$), and what small quadrotors experience ($\tau_d = O(10^{-1})\,s$ [18], corresponding to $\mathbb{S} \approx 1$). Important advantages of a few percent may be available to larger, faster vehicles: for eagles $\tau_d = O(1)\,s$ [19], corresponding to $\mathbb{S} \approx 10$, where $\mathbb{E}_{TW}/\mathbb{E}_{QF} \approx 0.92$, though the physics of eagles include lift and will require different calculations than those presented here. For neutrally buoyant aquatic organisms, $\mathbb{S} = 0$ and $\mathbb{Q} = 0$ so that the maximum possible benefit from path optimization is not bounded until basal metabolic costs are incorporated into the description.

The full information required to find globally optimized trajectories cannot be available to real vehicles, since turbulence cannot be predicted beyond a timescale set by $L/W$ [*e.g.* 20]. Future work includes evaluating the influences of the features of real vehicles and flows, including for instance the geometry and rotations of the vehicles, and the possibility that their mass changes as fuel is consumed. Real turbulence is of course three dimensional and covers a broader range of scales in nature than the Kraichnan model we used, which may allow vehicles to access energy on scales favorable to a given vehicle, or may interfere with their ability to do so on any scale. Motion in the vertical direction enables an exchange between kinetic and potential energy that is useful for energy extraction by gliders, and which we have excluded from our analysis. Tailwind Turbulence provides a benchmark for what can be gained purely from turbulent fluctuations, and a way to evaluate how turbulence affects the performance of vehicles (such as how performance depends on the organization of turbulence in a particular flow). How closely the bound can be approached in practice when limited information about the flow is available to a vehicle is yet to be seen.

**Acknowledgements**